\documentclass[aps,pre,twocolumn,groupedaddress,showpacs,showkeys]{revtex4}
\usepackage{epsfig}
\usepackage{psfig}
\usepackage{exscale}
\def\lsim{\raise0.3ex\hbox{$<$\kern-0.75em\raise-1.1ex\hbox{$\sim$}}}
\def\gsim{\raise0.3ex\hbox{$>$\kern-0.75em\raise-1.1ex\hbox{$\sim$}}}
\newcommand{\be}{\begin{equation}}
\newcommand{\ee}{\end{equation}}
\newcommand{\ba}{\begin{eqnarray}}
\newcommand{\ea}{\end{eqnarray}}

\def\spose#1{\hbox to 0pt{#1\hss}}
\def\ltapprox{\mathrel{\spose{\lower 3pt\hbox{$\mathchar"218$}}
 \raise 2.0pt\hbox{$\mathchar"13C$}}}
\def\gtapprox{\mathrel{\spose{\lower 3pt\hbox{$\mathchar"218$}}
 \raise 2.0pt\hbox{$\mathchar"13E$}}}
\def\phv{\vec \phi}

\def\NT{N_\tau}
\def\nt{\ifmmode\NT\else$\NT$\fi}
\def\NS{N_\sigma}
\def\ns{\ifmmode\NS\else$\NS$\fi}

\def\p{^\prime}

\def\n{\noindent}
\def\nn{\nonumber}

\begin{document}

\preprint{BI-TP 2003/13}
\preprint{hep-lat/0305019}
\title{Critical behaviour and scaling functions of the three-dimensional $O(6)$ model}
\author{S. Holtmann}
\author{T. Schulze}
\affiliation{Fakult\"at f\"ur Physik, Universit\"at Bielefeld, D-33615
  Bielefeld, Germany}

\date{\today}
\begin{abstract}
\n We numerically investigate the three-dimensional $O(6)$ model on $12^3$ to
$120^3$ lattices within the critical region at zero magnetic field, as well as
at finite magnetic field on the critical isotherm and for several fixed couplings in
the broken and the symmetric phase. We obtain from the Binder cumulant at vanishing
magnetic field the critical coupling $J_c=1.42865(3)$. The universal
value of the Binder cumulant at this point is $g_r(J_c)=-1.94456(10)$. At the
critical coupling, the critical exponents $\gamma=1.604(6)$,
$\beta=0.425(2)$ and $\nu=0.818(5)$ are determined from a finite-size-scaling
analysis. Furthermore, we verify predicted effects induced by massless Goldstone
modes in the broken phase. The results are well described by the
perturbative form of the model's equation of state. Our $O(6)$-result is
compared to the corresponding Ising, $O(2)$ and $O(4)$ scaling
functions. Finally, we study the finite-size-scaling behaviour of the
magnetisation on the pseudocritical line.
\end{abstract}

\maketitle

\noindent{\rule[-.3cm]{5cm}{.02cm}} \\

\section{Introduction}

\n The chiral phase transition of Quantumchromodynamics (QCD) is of great interest
for the understanding of the early universe and the physics of heavy ion
collisions. For two massless quark flavours it is supposed to be of second order
and to lie in the same universality class as the three-dimensional $O(4)$ spin
model \cite{Pisarski:ms}-\cite{Wilczek:1992sf}. If these assumptions
are valid, one can use the knowledge of the spin model to understand the critical
behaviour of the QCD phase transition.\\
It has been shown \cite{Aoki:1997xh,Aoki:1997nn} that the scaling behaviour in
lattice simulations with two light quark flavours is indeed comparable to the
universal infinite volume scaling function of the $O(4)$ spin model if one uses
Wilson fermions, although the Wilson fermion action has no chiral symmetry on
the lattice. For staggered fermions however the $O(4)$ scaling function does not
match the QCD data. For two flavours on the lattice the staggered fermion action
has a remaining $U(1)\times U(1)$ chiral symmetry. As this symmetry lies in 
the same universality class as the three-dimensional $O(2)$ spin model, the data
has also been compared to approximated infinite volume $O(2)$ data. But the
$O(2)$ scaling function matches even worse than the $O(4)$ function.\\
Since the lattice sizes used in QCD are rather small, it might be better to
compare the data to universal finite-size-scaling functions. We have indeed found
\cite{Engels:2001bq} that the finite-size-scaling functions on the pseudocritical
line of $O(2)$ and $O(4)$ are compatible with staggered QCD data.\\ 
It has turned out problematic to check QCD data for further critical behaviour
found in spin models, e.g. the goldstone effect. In QCD with fermions in
the fundamental representation the chiral and the deconfinement phase transitions
occur at the same temperature. This could change the properties of the chiral
phase transition, as additional degrees of freedom are released.\\
However in QCD with fermions in the adjoint representation (aQCD) the two phase
transitions are separated, so one can study them individually. Since the
left-handed and right-handed spinors are indistinguishable in the adjoint
representation, the chiral symmetry group is $SU(2N_f)$ and not $SU(N_f)_L
\times SU(N_f)_R$ as in the fundamental representation. Thus, for two flavours the
symmetry group is $SU(4)$, which is isomorph to $SO(6)$, a subgroup of
$O(6)$. The universality class of QCD with adjoint fermions therefore has to be
that of the $O(6)$ spin model. In this paper our results for the universal
properties of this model will be shown, especially the scaling functions, which
are needed for our forthcoming study of aQCD, continuing the work of Karsch and
L\"utgemeier \cite{Karsch:1998qj}.\\
The model we investigate is the standard $O(6)$-invariant nonlinear 
$\sigma$-model, which is defined as 
\be
\beta\,{\cal H}\;=\;-J \,\sum_{<x,y>} \phv_x\cdot \phv_y
         \;-\; {\vec H}\cdot\,\sum_{x} \phv_x \;.
\label{act}
\ee
Here $x$ and $y$ are the nearest-neighbour sites on a three-dimensional 
hypercubic lattice, $\phv_x$ is a $6$-component unit vector at site $x$ 
and $\vec H$ is the external magnetic field. The coupling constant $J$ is 
considered as inverse temperature, therefore $J=1/T$. An additional 
term $\sum_{x}[\phv_x^2 +\lambda (\phv_x^2-1)^2]$ is often used in the
Hamiltonian with $\lambda$ tuned to minimize leading order corrections to
scaling. It is not applied here, because the appropriate $\lambda$ value of
the $O(6)$ model has not been calculated yet, and such a calculation is beyond
the scope of this paper.

\n If $H=|{\vec H}|$ is non-zero we can decompose the spin vector $\phv_x$ into 
a longitudinal (parallel to the magnetic field ${\vec H}$) and a transverse 
component 
\be
\phv_x\; =\; \phi_x^{\parallel} {\vec{e}_H} + \phv_x^{\perp} ~ \quad
{\rm with}~~ {\vec{e}_H}= {\vec H}/H~. 
\ee
\n The order parameter of the system, the magnetisation $M$, is the
expectation value of the lattice average $\phi^{\parallel}$ of the longitudinal
spin component
\be
M \;=\; \langle\: \frac{1}{V}\sum_{x} \phi^{\parallel}_x\: \rangle\;
 =\; \langle \,  \phi^{\parallel}\, \rangle~,
\label{truem}
\ee
$V=L^3$ is the volume of the lattice with $L$ points per direction.

\n At zero magnetic field ($H=0$) there is no special direction and 
the lattice average of the spins 
\be
\phv\;=\; \!\frac{1}{V}\sum_{x}\phv_x\;~
\ee
will have a vanishing expectation value on all finite lattices, 
$\langle \: \phv \: \rangle = 0$. As an approximate order parameter for $M$ at
$H=0$ one can take \cite{Talapov:1996yh} 
\be
M \;\simeq \; \langle |\phv|\, \rangle~.
\label{magmod}
\ee
Nevertheless, we can use $\phv$ to define the susceptibilities and the Binder
cumulant by
\ba
\chi_v \!\!&\!=\!&\!\! V\langle \: \phv^2\: \rangle~,\label{chi}\\
\chi \!\!&\!=\!&\!\! V(\langle \: \phv^2\: \rangle \,-\,M^2)~,\label{chi_v}\\
g_r \!\!&\!=\!&\!\! {\langle \: (\phv^2)^2 \: \rangle \over
 \langle \: \phv^2 \: \rangle^2} -3~.
\label{gr}
\ea
In the following section we describe our simulations at zero magnetic 
field and estimate the critical coupling $J_c$ from the Binder cumulant, 
the magnetisation and the susceptibilities. In Section \ref{simuH0} the critical
exponents $\nu$, $\beta$ and $\gamma$ are determined. With simulations
at $H>0$ in Section \ref{simuH} we investigate the behaviour of the model on the
critical line, in the broken phase and in the symmetric phase. Finally, the
resulting data is used in Section \ref{ScalingFct} to generate the
infinite volume scaling function of the magnetisation. Using this data the
infinite volume scaling function of the susceptibility and the position of the
pseudocritical line are derived in Section \ref{PCL}. A summary and our
conclusions are given in Section \ref{section:conclusion}. 

\section{Simulations at $H=0$}
\label{simuH0}

\begin{table}[b]
\begin{tabular}{|r|c|c|c|c|c|c|}
\hline
$L$ & $J$-range & $N_J$ & $N_{meas} [1000]$   
& \multicolumn{1}{|c|}{$\tau_{int}$} \\
\hline
   12 & 1.42830 - 1.42900 & 25 & 100 - 200 & $\lsim$ 3 \\
   16 & 1.42840 - 1.42880 & 18 & 100 - 200 & $\lsim$ 4 \\
   20 & 1.42840 - 1.42885 & 19 & 100 - 200 & $\lsim$ 6 \\
   24 & 1.42835 - 1.42885 & 19 & 100 - 200 & $\lsim$ 6 \\
   30 & 1.42840 - 1.42885 & 18 & 100 - 200 & $\lsim$ 6 \\
   36 & 1.42840 - 1.42880 & 17 &    100    & $\lsim$ 5 \\
   48 & 1.42840 - 1.42880 & 17 &    100    & $\lsim$ 6 \\
   60 & 1.42840 - 1.42880 & 16 &    80     & $\lsim$ 8 \\
   72 & 1.42840 - 1.42880 & 16 &    80     & $\lsim$ 9 \\
\hline
\end{tabular}
\caption{Survey of the Monte Carlo simulations at $H=0$ for different
lattices. Here $N_J$ is the number of different couplings performed in the
appropriate $J$-range; $\tau_{int}$ is the integrated autocorrelation time for
the energy and $N_{meas}$ the number of measurements per coupling in units of 1000.}
\label{tab:survey}
\end{table}

\n All our simulations were done on three-di\-mensional cubic lattices with periodic
boundary conditions. We used Wolff's single cluster algorithm as we did in our
previous papers (Refs.\ \cite{Engels:2001bq} and
\cite{Engels:1999wf}-\cite{Cucchieri:2002hu}). The $H=0$ data  were taken from
lattices with linear extensions $L=12,16,20,24,30,36,48,60$ and 72. Between the
measurements we performed 300-600 cluster updates to reduce the integrated
autocorrelation time $\tau_{int}$ for the energy.\\
Butera and Comi \cite{Butera:1998rk} determined the critical point of the $O(6)$
spin model using a high temperature (HT) expansion as $J_c=1.42895(6)$. Therefore
we generally scanned the range from $J=1.3$ up to $J=1.55$ on smaller
lattices with careful regard to the critical region close to the $J_c$-value found in
\cite{Butera:1998rk} for all lattices. This data was then further analysed using
the reweighting method. More details of the simulations near the critical
point are presented in Table \ref{tab:survey}.

\subsection{The Critical Point $T_c$}
\label{TCP}
\begin{figure}[ht]
   \epsfig{bbllx=83,bblly=214,bburx=496,bbury=587,
       file=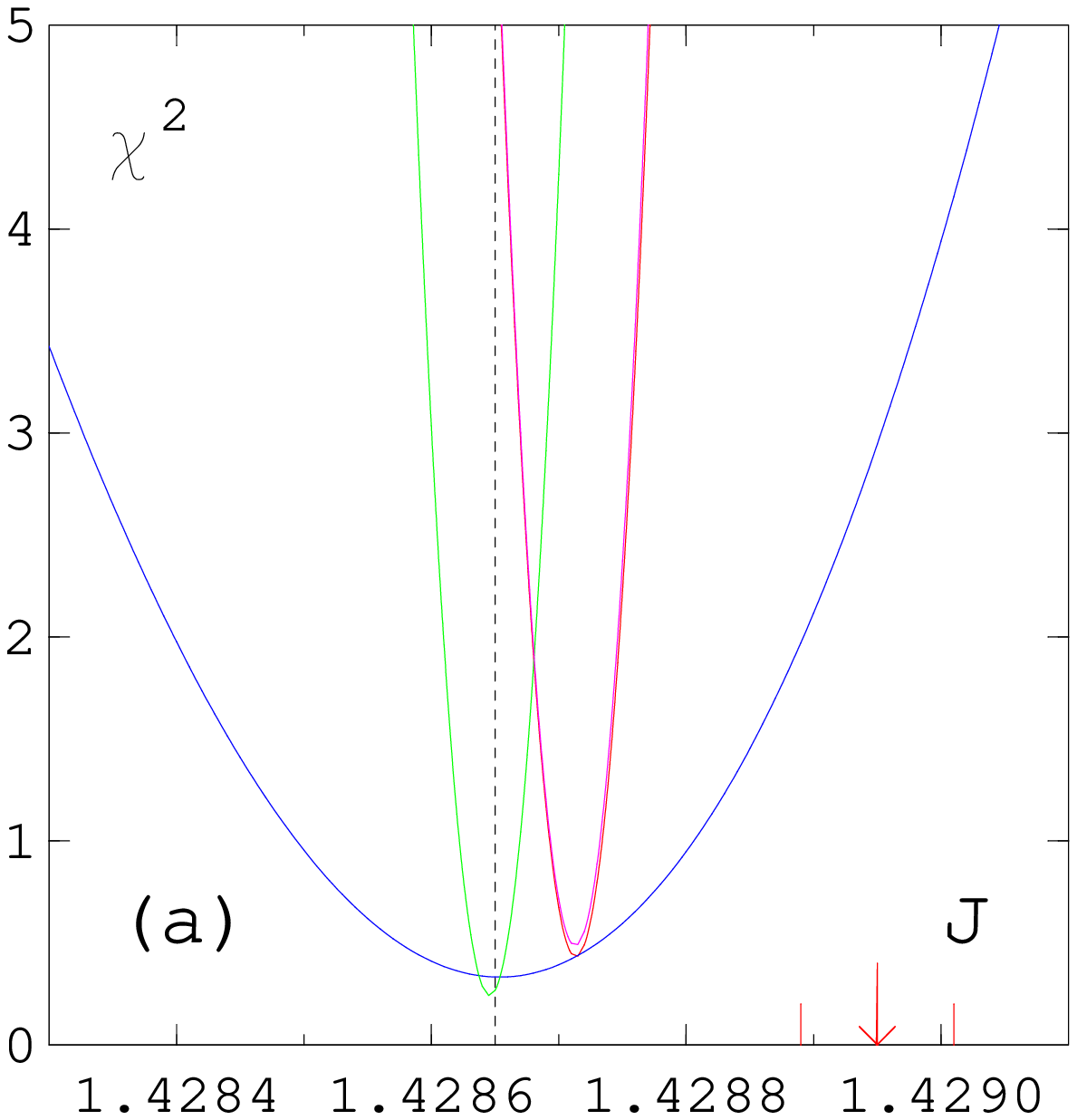, width=70mm}
   \epsfig{bbllx=83,bblly=250,bburx=496,bbury=587,
       file=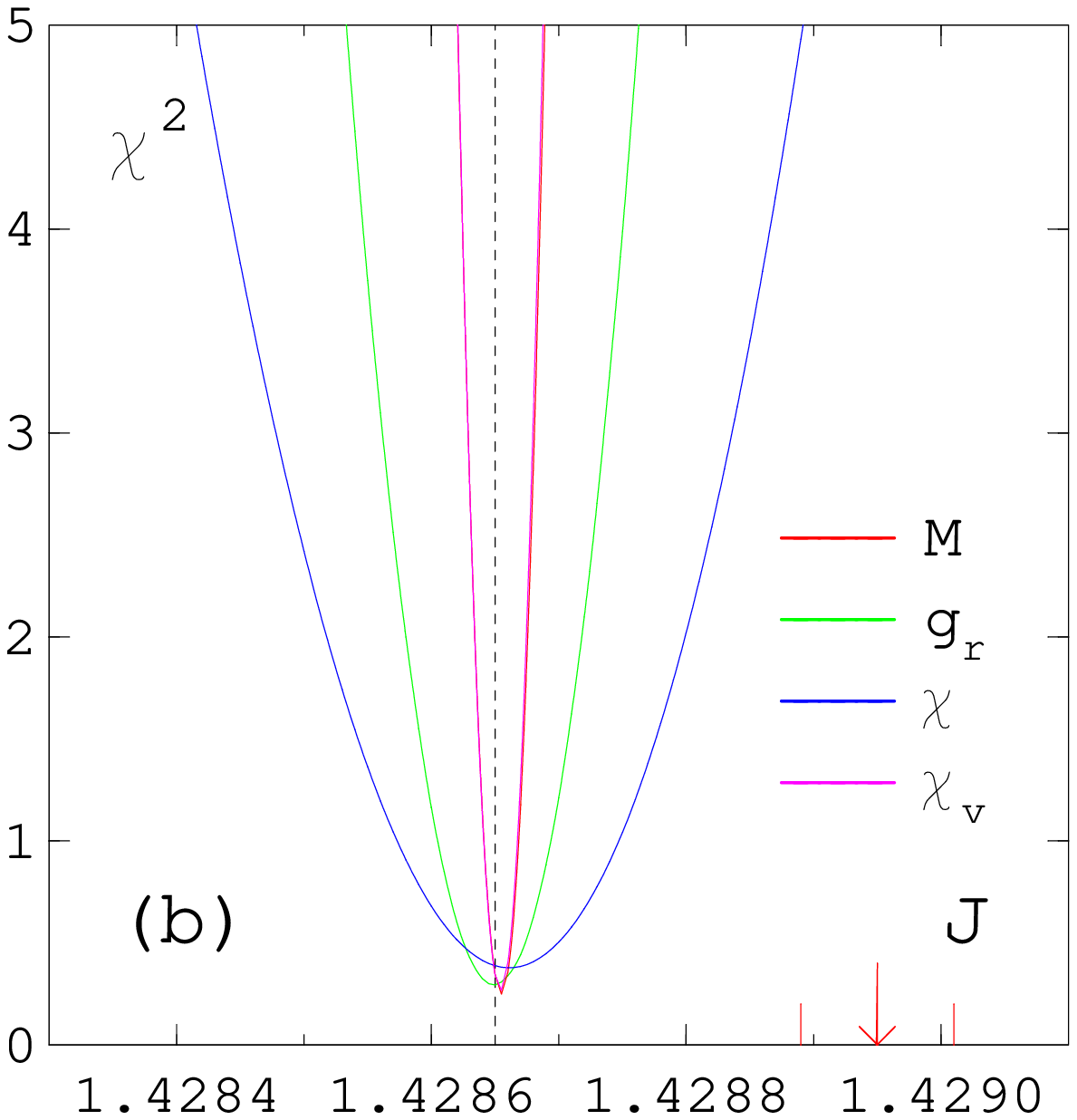, width=70mm}
\caption{The minimal $\chi^2$ per degree of freedom for fits according to
   equation (\ref{obsTc}) for $M$ (red), $\chi_v$ (pink), $\chi$
   (blue) and $g_r$ (green). Figure (a) shows the
   results without a correcting term $\propto L^{-\omega}$, whereas figure (b)
   uses $\omega=0.5$. The dotted black lines show our previous value of $J_c$, the
   arrows the result of \cite{Butera:1998rk} with its error bars. The curves of
   $M$ and $\chi_v$ lie on top of each other.}
\label{chi2}
\end{figure}

\n Obviously any determination of critical values as well as the definition of
the reduced temperature
\be
t\;=\;{{T-T_c}\over T_c}
\ee
relies on the exact location of the critical point. Since there is no result from
numerical studies, we check the aforementioned value of Butera and Comi
\begin{figure}[ht]
   \epsfig{bbllx=83,bblly=214,bburx=496,bbury=587,
       file=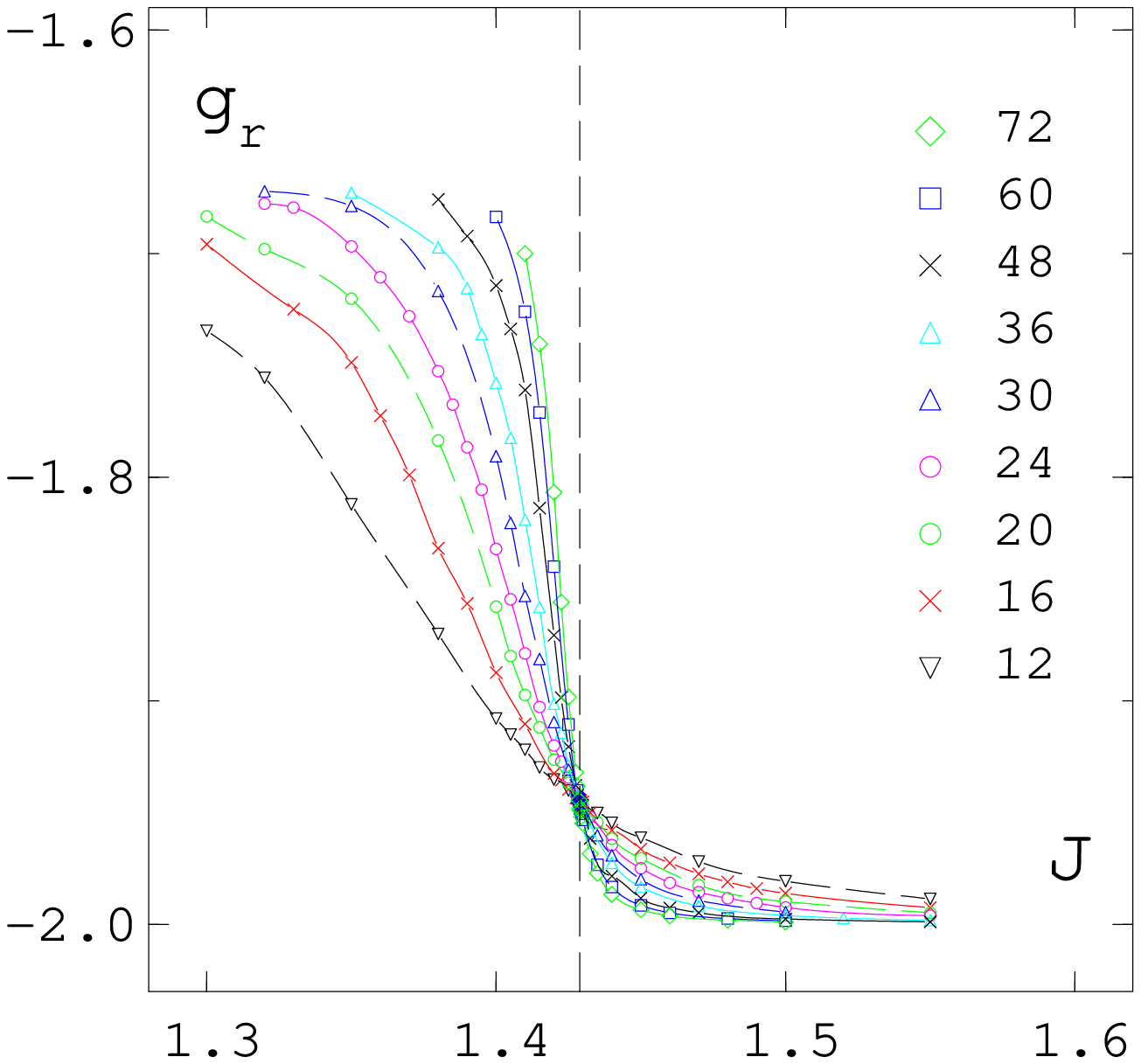, width=70mm}
   \epsfig{bbllx=83,bblly=250,bburx=496,bbury=587,
       file=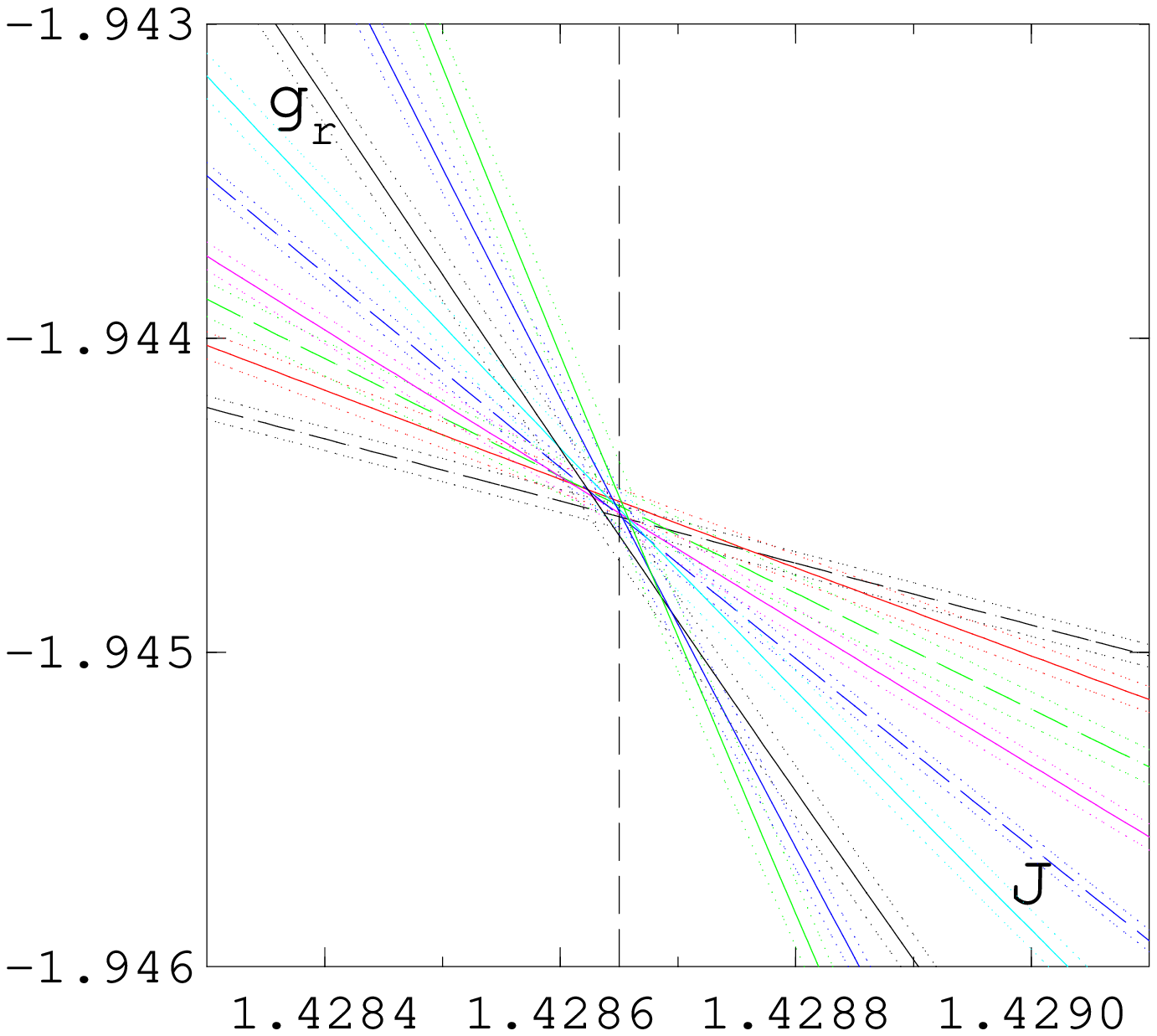, width=70mm}
\caption{(a) The Binder cumulant $g_r$ from Eq.\ (\ref{gr}) as a function of the
       coupling $J$. The points are connected by splines to guide the eye. With
       increasing lattice size $L= 12,16,20,24,30,36,48,60$ and $72$, the slope
       of the respective curve in the critical region increases. The vertical
       dashed line denotes our final result for $J_c$. (b) is an enlargement of
       (a) near the critical point. The dotted lines accompanying the
       solid/dashed lines show the jackknife error corridors.}
\label{fig:gr}
\end{figure}
\n first. We determine $T_c$ by studying the Binder cumulant $g_r$, which is a
finite-size-scaling function
\be
g_r \;=\; Q_g (t L^{1/\nu}, L^{-\omega})~.
\label{grscl} 
\ee
The function $Q_g$ depends on the thermal scaling field and on possible
irrelevant scaling fields. In this case only the leading irrelevant scaling field
proportional to $L^{-\omega}$ is specified, with an unknown $\omega>0$. 
Therefore, at the critical point ($t=0$) $g_r$ ought to be independent of $L$ apart from 
corrections due to these irrelevant scaling fields. Fig.\ \ref{fig:gr} (a) shows
our results for $g_r$. On the scale of Fig.\ \ref{fig:gr} (a) we observe no
deviation from the scaling hypothesis. After a blow-up of the close
vicinity of the critical point, as shown in Fig.\ \ref{fig:gr} (b), one sees
that the intersection points $J_{ip}$ between the curves of different 
lattices are not coinciding perfectly at one $J$. These minor corrections to
scaling have to be considered. By expanding the scaling function $Q_g$ to lowest
order in both variables one gets for the intersection point $J_{ip}$ of two lattices with
sizes $L$ and $L\p=bL$ 
\be
J_{ip}(L,b)\;=\;J_c\,+\,c_1\,s(L,b)
\ee
with
\be
s(L,b)\;=\;{ 1 -b^{-\omega} \over b^{1/\nu} -1} L^{-\omega -1/\nu}~.
\ee
To have an unbiased estimate of $J_c$ we choose Binder's approximation \cite{Binder}
\be
{1 \over J_{ip}}\;=\;{1 \over J_c}\,+\,{c_2 \over \ln b}~,
\label{shift}
\ee
which can be used without knowing the values of $\nu$ and $\omega$. In Fig.\
\begin{figure}[t]
   \epsfig{bbllx=83,bblly=264,bburx=496,bbury=587,
       file=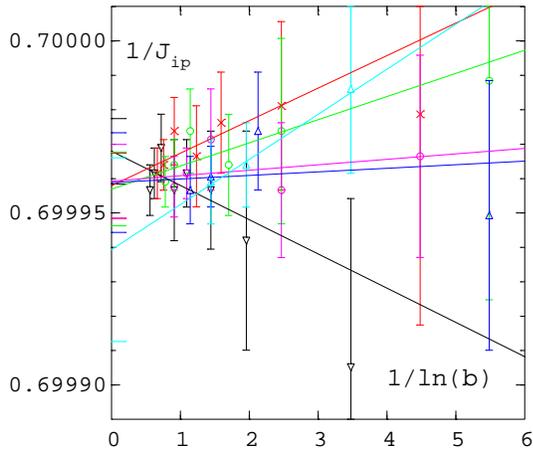, width=70mm}
\caption{\label{fig:jc}The coupling $J_{ip}$ at the intersection point of $g_r(L)$ and 
$g_r(bL)$ for various $L$ and $b$ as a function of $1/\ln b$, Eq.\
(\ref{shift}). The lines are linear fits for the lattices $L=12$ (black),
$16$ (red), $20$ (green), $24$ (pink), $30$
(dark blue) and $36$ (light blue) through the intersection points with
all larger lattices. The errors of the endpoints are drawn on the ordinate.}
\end{figure}
\n \ref{fig:jc} the $1/J_{ip}$ values of the intersection points from the
lattices $L=12,16,20,24,30,36$ with all other larger lattices are plotted as a
function of the variable $1/\ln b$ of Eq.\ (\ref{shift}). Linear fits should
lead to the same value $1/J_c$ within the errors. Fitting the results to a
constant value and varying the used $L$-values we find
\be
{1 \over J_c}\;=\;0.699960(14),
\ee
which is equivalent to
\be
J_c\;= \; 1.42865(3).
\label{JcResult}
\ee
This result agrees in the first four digits with the result $J_c=1.42895(6)$ of
Butera and Comi \cite{Butera:1998rk}. There is a slight difference in the
last two digits. As this difference is larger than the corresponding errors, we
check our result with the $\chi^2$-method \cite{Engels:1995em} described in the
following.\\
Let us review the general form of the scaling relations for different
observables $\cal O$
\be
{\cal O}\;=\;L^{\rho /\nu}\,Q_{\cal O}\,(tL^{1/\nu},L^{-\omega}),
\ee
where we only take the largest irrelevant exponent into account. Here
$\cal O$ is $M$, $\chi$ or $g_r$ with $\rho=-\beta,\gamma$
and $0$ respectively. Expanding the function $Q_{\cal O}$ to first order in the
variables we find
\ba
{\cal O}\!\!\!\!\!\!\!&=&\!\!\!\!\!\!\!L^{\rho
  /\nu}\,(c_0\,+\,(c_1\,+c_2L^{-\omega})\,tL^{1/\nu}\nn\\
&&{}\!\!\!\!\!\!\!+\,c_3L^{-\omega}), 
\ea
which reduces to
\be
{\cal O}\;=\;L^{\rho /\nu}\,(c_0\,+\,c_3L^{-\omega})
\label{obsTc}
\ee
at the critical point $t=0$. Therefore at the critical coupling a fit with
equation (\ref{obsTc}) has the minimal $\chi^2$. Since we do not know the
influence of $c_3L^{-\omega}$ we started without this correction term leaving
$L=12$ out. Fig.\ \ref{chi2} (a) shows the result. we find a deviation from our
preliminary result in case of the magnetisation $M$ and the susceptibility
$\chi_v$. The minima from the Binder cumulant and the susceptibility $\chi$
however coincide at $J\approx 1.42865$.\\
\n We thereafter made fits with the correction term in the range $\omega = 0.5 -
1.5$. The minimal $\chi^2$ and a perfect agreement of $J_c$ for all observables is found at
$\omega = 0.5$. This result is plotted in Fig.\ \ref{chi2} (b). $\chi^2/d.o.f.$
increases with $\omega$ and shifts $J_c$ in case of $M$ and $\chi_v$ to smaller
couplings, while the position calculated from $\chi$ increases and $J_c$ from
$g_r$ remains nearly constant. Since the fits get worse we can exclude
$\omega$-values larger than $0.8$. The positions of $J_c$ for
$0.5\leq\omega\leq 0.8$ coincide within the error bars of Eq. (\ref{JcResult}).\\
At the critical point the Binder cumulant has the form
\be
g_r(L)\;=\;g_r(J_c)\,+\,c_3L^{-\omega},
\ee
with the universal value $g_r(J_c)$ and a small correction term
$c_3L^{-\omega}$. For fits with different $\omega$ we find $g_r$
\be
g_r(J_c)\;=\;-1.94456(10).
\ee
The quality of the fits does not change much ($\chi^2/d.o.f.\approx 0.2-0.3$)
with different $\omega$ so a better estimate of $\omega$ is still not possible.
\subsection{The critical exponents}
\begin{figure}
   \epsfig{bbllx=83,bblly=254,bburx=496,bbury=587,
       file=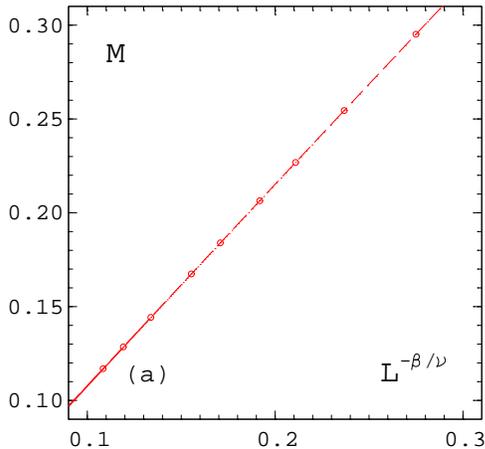, width=70mm}
\caption{The magnetisation $M$ as a function of the lattice extension $L$ at the critical
  point $T_c$ and $H=0$. The dashed line in is a fit to the ansatz (\ref{ML})
  with $\omega=0.5$.}
\label{FSjcmag}
\end{figure}
\begin{figure}
   \epsfig{bbllx=83,bblly=254,bburx=496,bbury=587,
       file=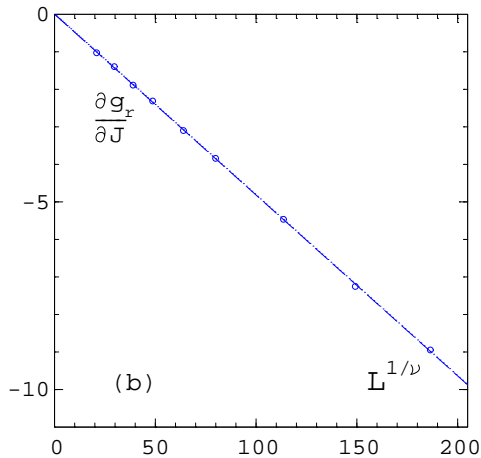, width=70mm}
\caption{The derivative of the binder cumulant $\partial g_r/\partial J$ as a
       function of the lattice extension $L$ at the critical point $T_c$ and
       $H=0$. The dashed line is a fit to the ansatz (\ref{gr_deri}) with
       $d_3=0$ since the corrections are negligible.}
\label{FSjcgr}
\end{figure}
\n Since we now know the critical coupling we can study the finite-size
behaviour of several observables with Eq.\ (\ref{obsTc}). These can be extracted
from our reweighted data at $T_c$. The studied scaling relations are
\ba
\label{ML}
M\!\!&\!=\!&\!\!L^{-{\beta /\nu}}\,(a_0\,+\,a_3L^{-\omega}),\\
\chi \!\!&\!=\!&\!\!L^{\gamma /\nu}\,(b_0\,+\,b_3L^{-\omega})
\ea
with $\omega \in [0.5;1.0[$ and the exponents $-\beta / \nu$ and
$\gamma / \nu$ as free parameters. Since these two ratios are connected
by the hyperscaling relation
\be
{\gamma \over \nu}\;=\;3\,-\,{2\beta \over \nu},
\ee
it is necessary to study a further observable, for example the derivative of
$g_r$, which is given at $T_c$ by
\be
{\partial g_r \over \partial J}\;=\;L^{1 /\nu}\,(d_0\,+\,d_3L^{-\omega}).
\label{gr_deri}
\ee
In this way two independent exponents (e.g. $\beta$ and $\nu$) can be
estimated. We fit all observables in the range $L=12-72$. From
Eq. (\ref{ML}) we obtain
\be
{\beta \over \nu}\;=\;0.519(2),
\label{betaovernu}
\ee
in which the error also includes an $\omega$-variation in $[0.5, 1.0[$. Here a
\begin{table*}
\begin{tabular}{|c|c|c||c|c|c|c|c|}
\hline
 Source & & & exponent & this work &
 \multicolumn{1}{|c|}{\cite{Butera:1998rk}}&\cite{Antonenko:1998es}&
 {\cite{Gracey:1991zq,Graceypc}}\\
\hline
$\partial g_r/\partial J$ & $1/\nu      $ & 1.223(5) & $\nu   $ & 0.818(5) &
 0.819(3) & 0.790 & 0.819 \\
   $M$                    & $\beta /\nu $ & 0.519(2) & $\beta $ & 0.425(2) &
 0.424(5) & 0.407 & 0.424 \\
   $\chi$                 & $\gamma /\nu$ & 1.961(3) & $\gamma$ & 1.604(6) &
 1.608(4) & 1.556 & 1.609 \\
\hline
\end{tabular}
\caption{The critical exponents for the $O(6)$ model estimated in this work
 compared to the theoretical work of Butera and Comi \cite{Butera:1998rk}, 
 Antonenko and Sokolov \cite{Antonenko:1998es} and Gracey {\cite{Gracey:1991zq,Graceypc}}.}
\label{critexpo}
\end{table*}
larger $\omega$ shifts $\beta/\nu$ to a smaller value at nearly constant
$\chi^2/d.o.f.\approx 0.4$. Fig.\ \ref{FSjcmag} shows the result with $\omega=0.5$.\\
Our $\chi$-fits yield
\be
{\gamma \over \nu}\;=\;1.961(3).
\label{gammaovernu}
\ee
Our results of $\beta/\nu$ and $\gamma/\nu$ are tested with the hyperscaling relation 
\be
2{\beta \over \nu}\,+\,{\gamma \over \nu}\;=\;d,
\ee
with $d=3$ being the dimension of the model. The left hand side of this equation
is $2.999(5)$, correct within the error.\\ 
Finally we analyse the derivative $\partial g_r/\partial J$ of the Binder
cumulant at $T_c$. This observable is directly calculated from the spline
connection of our reweighted data in the neighbourhood of the critical
point. The errors are obtained with the jackknife method, which seems to
underestimate the errors, so we therefore assume the largest error of the
different $L$-values for each lattice. Our fit to ansatz (\ref{gr_deri}) without
corrections to scaling ($\chi^2/d.o.f.\approx 0.3$) is shown in Fig.\
\ref{FSjcgr}. For $d_3=0$ we find
\be
{1 \over \nu}\;=\;1.223(5).
\label{1overnu}
\ee

\n The final results of the critical exponents are summarized in Table \ref
{critexpo}. $\beta$ and $\gamma$ are calculated with the result of $\nu$ and the
ratios (\ref{betaovernu}) and (\ref{gammaovernu}). The three last columns of the
table show the results from \cite{Butera:1998rk}, \cite{Antonenko:1998es} and
\cite{Gracey:1991zq,Graceypc}. Butera and Comi as well as Gracey are in good
agreement with our values, but the results of Antonenko and Sokolov are farther away.\\
In the following Sections we use the fixed critical exponents $\beta=0.425$ and
$\nu=0.818$. The remaining critical exponents are calculated by the respective
hyperscaling relations between the critical exponents. For $\omega$ we will use
the value $\omega=0.5$, which seems to be the best estimate in all investigations.

\section{Simulations at $H>0$}
\label{simuH}
\n The magnetisation $M$ is now calculated from equation (\ref{truem}). A
transversal and a longitudinal susceptibility can be defined as
\ba
\chi_L\;&=&\; V(\langle \: M^2\: \rangle \,-\,M^2)~,\label{chi_L}\\
\chi_T\;&=&\; V\langle \: (\phv^{\perp})^2 \: \rangle~.\label{chi_T}
\ea
We simulated at several constant $J$-values and increasing magnetic field,
starting at $H=0.00025$. The used lattice sizes were $L=24,36,48,72,96$ and
$120$. Around $20,000$ measurements were performed in the $(J,H)$-regions we used
for our fits. The only exception was the data of $L=120$, where we
performed $10,000$ measurements at $J_c$ and $5,000$ measurements at all other
$J$-values. The integrated autocorrelation time for the energy and the
magnetisation is strongly dependent on the used $J$-values. At $J_c$ and
$J>J_c$ we increased the number of cluster updates between two measurements to
have autocorrelation times $\tau_{int}\lsim 6$.\\
In the symmetric phase ($J<J_c$) the situation is different. While the
measurements of the magnetisation are less correlated with
$\tau_{int}(M)\lsim 4$, the correlation of the energy increases rapidly with
decreasing $H$ and $J$. It reaches values of $\tau_{int}(E)\lsim 30$ for the
larger lattices. 

\subsection{The critical isotherm}
\label{z=0}
\n At the critical point the critical scaling of the magnetisation is given by
\be
M(T_c,H) \;=\;d_c H^{1/\delta}\,(1\,+\,d_c^1H^{\omega\nu_c})~,
\label{MatTc}
\ee
where non-analytic corrections from the leading irrelevant scaling field are
taken into account. They are not negligible in our model. The critical
exponents $\delta$ and $\nu_c$ are known from the hyperscaling relations and
only depend on the ratio $\beta/\nu=0.519(2)$: 
\ba
\delta\;&=&\;3{\nu \over \beta}-1\;=\;4.780(22),\\
\nu_c\;&=&\;{\nu \over \beta\delta}\;=\;0.4031(24).
\ea
\begin{figure}[ht]
   \epsfig{bbllx=83,bblly=264,bburx=496,bbury=587,
       file=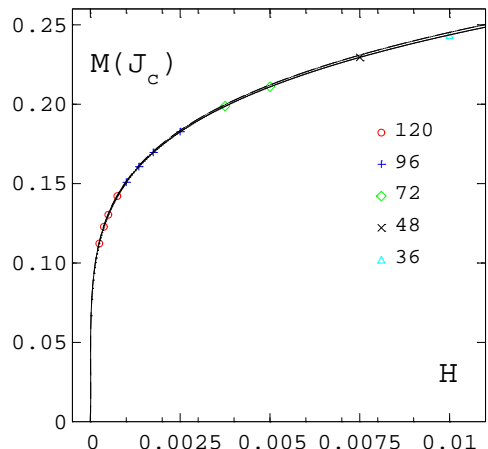, width=70mm}
\caption{The magnetisation at the critical coupling as a function of $H$. The
       solid line is the fit to the ansatz (\ref{MatTc}), while the dashed line
       is the leading term.}
\label{magJc}
\end{figure}
\begin{figure}[ht]
   \epsfig{bbllx=83,bblly=264,bburx=496,bbury=587,
       file=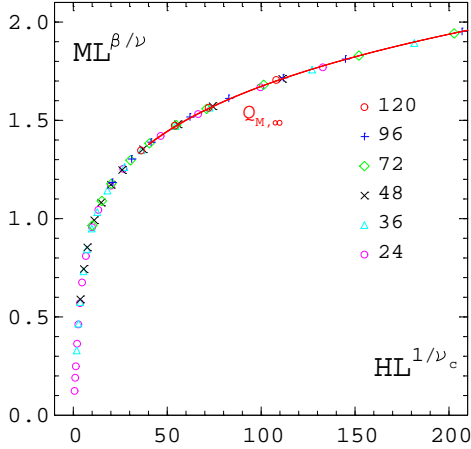, width=70mm}
\caption{The finite-size-scaling function $Q_{M,\infty}$ on the critical line, Eq.\
       (\ref{FSSQM}). The solid red line indicates the asymptotic function
       $Q_{0,\infty}$ for $z'\gsim 40$.}
\label{FSS1}
\end{figure}
\n In order to exclude finite-size effects we carry out a reweighting analysis for
all lattices and fit the result from the largest lattice to approximate the
value of $V \to \infty$. This is done for the interval $H\in [0.00075;0.04]$
and we find
\be
d_c=0.642(1).
\ee
\n Our result is plotted in Fig.\ \ref{magJc}. There are minimal
negative corrections. If one treats $\delta$ as a free parameter the result
$\delta=4.79(1)$ agrees with our first estimate.\\
The finite-size-scaling function for the magnetisation is
\ba
M(T,H,L)\!\!\!\!\!\!\!\!&=&\!\!\!\!\!\!\!\!L^{-\beta/\nu}\cdot\nn\\
&&\!\!\!\!\!\!\!\!\Phi ( tL^{1/\nu}, HL^{1/\nu_c}, L^{-\omega}).
\label{fssm}
\ea
The scaling function $\Phi$ can be expanded in $L^{-\omega}$ to
\ba
M(T,H,L)\!\!\!\!\!\!\!\! &=&\!\!\!\!\!\!\!\! L^{-\beta/\nu}\, \Phi_0 (
tL^{1/\nu}, HL^{1/\nu_c})\nn\\
&&{}\!\!\!\!\!\!\!\!+ \dots\;.
\label{mexpan}
\ea
At $T_c$ the leading part is now given by
\be
M(T_c,H,L)\;=\;L^{-\beta / \nu}\,Q_M(z')
\label{FSSQM}
\ee
with the universal scaling function $Q_M(z')$ and the argument
$z'=HL^{1/\nu_c}$. The results of all lattices are shown in Fig.\
\ref{FSS1}. The data points scale very well and the influence of corrections to
scaling is small. In the limit $z'\to\infty$ we expect the asymptotic behaviour
\be
Q_{M,\infty}(z')\;=\;d_c\,z'^{1/\delta},
\ee
which is observable for $z'\gsim 40$. This way one checks the value of
$d_c$ with a fit of reweighted $z'$-data of the larger lattices
$L=72,96,120$. We find $d_c=0.642(1)$, which agrees perfectly with our first
value of $d_c$.
\subsection{Numerical results at $T\neq T_c$}
\begin{figure}[t]
   \epsfig{bbllx=83,bblly=264,bburx=496,bbury=587,
       file=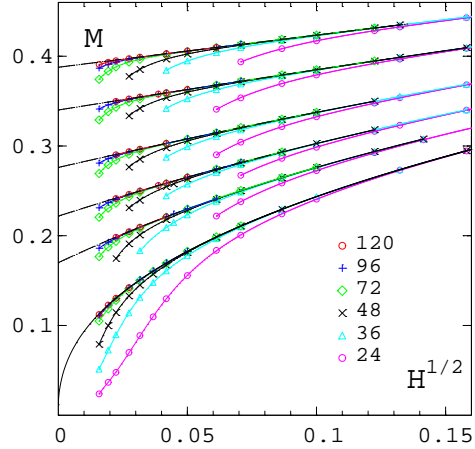, width=70mm}
\vspace{0.5cm}
\caption{The magnetisation in the broken phase as a function of $\sqrt{H}$ for the
       couplings $J=1.6$, $1.55,1.50,1.47,1.45$ and $J_c$ and different lattices,
       starting with the largest $J$-value at the top. The solid lines represent
       interpolations from a reweighting analysis of the data. The dashed lines
       are the fits to ansatz (\ref{quamag}) while the dotted line is the fit of
       equation (\ref{MatTc}) at $J_c$.}
\label{mag3}
\end{figure}
\n Let us review some perturbative predictions for the magnetisation and the
susceptibilities. The continuous $O(6)$ symmetry of our spin model gives rise
to spin waves, which are slowly varying (long-wavelength) spin configurations
with energies arbitrarily close to the ground-state energy. In $d>2$ they are
massless Goldstone modes associated with the spontaneous breaking of the
rotational symmetry for temperatures below the critical value $T_c$
\cite{spin}. For $T<T_c$ the system is in a broken phase, i.e.\ the
magnetisation $M(T,H)$ attains a finite value $M(T,0)$ at $H=0$.\\
The transverse susceptibility has the form
\be
\chi_T\;=\;\frac{M(T,H)}{H}
\label{chiT}
\ee
for all $H$ and $T$. This relation is a direct consequence of the $O(6)$
invariance of the zero-field free energy and can be derived as a Ward identity
\cite{Brezin}.\\
\begin{table*}
  \begin{tabular}{|c|ccc|c|c|}
    \hline
     $J=1/T$ & $M(T,0)$ &$c_1(T)$ &$c_2(T)$  &$10^{4}\cdot H$&$\chi^2/dof$\\ \hline
      1.45   &0.1701(03)&1.339(15)&-2.86(18) &   9-25        &    0.45    \\ 
      1.47   &0.2219(02)&0.924(02)&-1.138(14)&  12-74        &    0.56    \\
      1.50   &0.2761(01)&0.659(01)&-0.436(08)&  10-93        &    0.27    \\
      1.55   &0.3401(01)&0.463(01)&-0.141(02)&  10-163       &    0.78    \\
      1.60   &0.3878(01)&0.363(01)&-0.047(01)&  19-175       &    0.49    \\ \hline
  \end{tabular}
\caption{Parameters of the fit of $M(T<T_c,0)$ to the ansatz (\ref{quamag}). The
     fifth column is the used fit range.}
\label{tabM0}
\end{table*}
\n The longitudinal susceptibility diverges on the coexistence curve for $2<d \leq
4$ \cite{goldstone,WZ}. The leading terms in the perturbative expansion for three
dimensions are
\be
\chi_L(T<T_c,H)\;=\; b_0(T)\,H^{-1/2}\,+\,c_2(T)~.
\label{chiL}
\ee
\begin{figure}[ht]
   \epsfig{bbllx=83,bblly=254,bburx=496,bbury=587,
       file=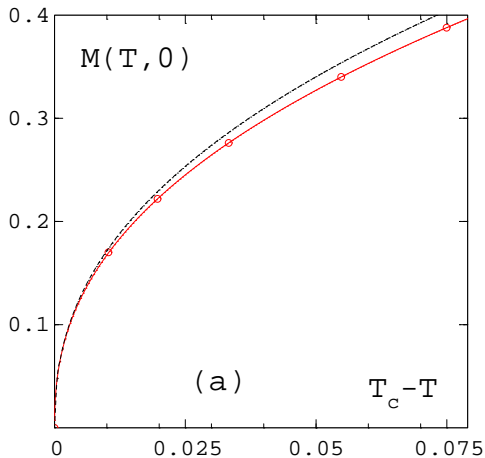, width=70mm}
\caption{Magnetisation M(T,0) on the coexistence curve as a function of
  $(T_c-T)$. The red line is the fit to ansatz (\ref{cocrit}) while the black dashed
  line is the leading part.}
\label{magT0}
\end{figure}
\n Since the susceptibility is the derivative of the magnetisation with respect to
$H$ we find for the magnetisation
\ba
M(T<T_c,H)\!\!\!\!\!\!\!\!&=&\!\!\!\!\!\!\!\!M(T,0)\,+\,c_1(T)\,H^{1/2}\nn\\
\!\!\!\!\!\!\!\!&&\!\!\!\!\!\!\!\!+\,c_2(T)\,H
\label{quamag}
\ea
near the coexistence curve. Fig.\ \ref{mag3} shows our results of the magnetisation in the
broken phase and the corresponding extrapolations to $M(T,0)$ in the
thermodynamic limit ($V\to\infty$). The numbers of the parameters are presented
in Table \ref{tabM0}. The $H$-extension of the regions, where the predicted
Goldstone behaviour is found, increases with $J$, while finite-size effects
become larger at small $H$ and larger $J$ ($L\gsim 160$ would be necessary for
finite-size independent data).\\
We fitted the values of $M(T,0)$ to the form
\ba
M(T\ltapprox T_c,0)\!\!\!\!\!\!\!\!&=&\!\!\!\!\!\!\!\!B\,(T_c-T)^{\beta}[1\,+\,b_1\,(T_c-T)^{\omega \nu}\nn\\
\!\!\!\!\!\!\!\!&&\!\!\!\!\!\!\!\!{}+\,b_2\,(T_c-T)]
\label{cocrit}
\ea
with fixed values $\beta=0.425$, $\omega \nu=0.409$ and the result
\be
B=1.22(1),
\ee
$b_1=-0.184(49)$ and $b_2=0.31(13)$. The error of $B$ also
\begin{figure}[ht]
   \epsfig{bbllx=83,bblly=254,bburx=496,bbury=587,
       file=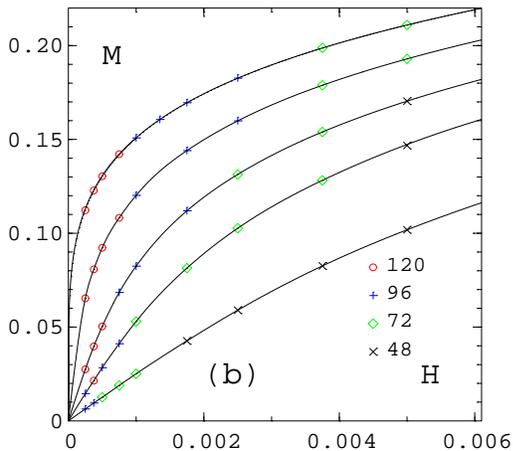, width=70mm}
\caption{The magnetisation in the symmetric phase 
  ($T\ge T_c$) as a function of $H$, starting from the top with fixed
  $J=J_c,1.42,1.41,1.40$ and $1.38$ and different $L$-values. The lines are
  spline connections between the data points.}
\label{magHOT}
\end{figure}
\n includes the slight uncertainty in the value of $\omega \nu$. Our final result
of $M(T,0)$ and the difference to the leading term are plotted in Fig.\
\ref{magT0}.\\
Since one of the main aims of this work is the determination of the magnetic
equation of state in Section \ref{ScalingFct}, we also simulated in the
high temperature phase. Again we use data of the largest lattices as an
approximation for the infinite volume value. The result is plotted in Fig.\
\ref{magHOT}. From a Taylor expansion we expect
\be
M(T > T_c,H)\;\propto\;H.
\label{MpropH}
\ee
at small $H$. With increasing temperature the $H$-interval with this behaviour
increases, as one can see in Fig.\ \ref{magHOT}.

\section{The Scaling function}
\label{ScalingFct}
\begin{figure}[ht]
   \epsfig{bbllx=83,bblly=260,bburx=496,bbury=587,
       file=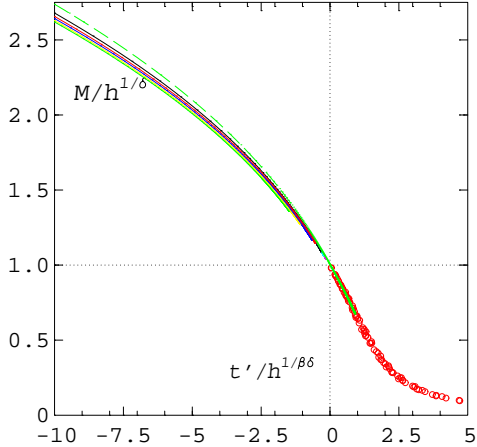, width=70mm}
\caption{The function $M/h^{1/\delta}$. The solid lines in the broken phase are
       the reweighted results for $M/h^{1/\delta}$ at
       $J=1.60,1.55,1.50,1.47,1.45,1.445$ and $1.44$, from the bottom to the
       top. They are extrapolated with Equation (\ref{fG_corr}) to $f_G$ (dashed
       line). The circles are single data points in the symmetric phase.} 
\label{fig:scafun}
\end{figure}
\n The critical behaviour of the magnetisation in the vicinity of $T_c$ is
described by the general Widom-Griffiths form \cite{Griffiths}
\be
y\;=\;f(x)
\label{eqstate}
\ee
with
\be
y \equiv h/M^{\delta}, \quad x \equiv t'/M^{1/\beta},
\label{xy}
\ee
where the variable $t'$ is proportional to $(T-T_c)$ and $h$ proportional to $H$. A common
normalisation of the function $f(x)$ is
\be
f(0) = 1, \quad f(-1) = 0.
\label{normal}
\ee
The variables $t'$ and $h$ are the conveniently normalized reduced temperature
$t'=(T-T_c)/T_0$ with $T_0=B^{-1/\beta}=0.626(12)$ and the reduced magnetic
field $h=H/H_0$ using $H_0=d_c^{-\delta}=8.32(6)$.
The function $f(x)$ is universal and was derived from the $\epsilon$-expansion
($\epsilon=4-d$) to order $\epsilon^2$ \cite{Brezin}. In the limit $x\to -1$,
i.e.\ at $T<T_c$ and close to $H=0$ the result was inverted to give $x+1$ as
a double expansion in powers of $y$ and $y^{d/2 - 1}$ \cite{WZ}
\ba
x+1\!\!\!\!\!\!\!\!&=&\!\!\!\!\!\!\!\! {\widetilde c_1} y \,+\, {\widetilde c_2}
y^{d/2 - 1}\,+\,{\widetilde d_1} y^2\nn\\
&&{}\!\!\!\!\!\!\!\!+\, {\widetilde d_2} y^{d/2}\,+\,{\widetilde d_3} y^{d-2}
\,+\, \ldots \;.
\label{f_inv}
\ea
The coefficients ${\widetilde c_1}$, ${\widetilde c_2}$ and ${\widetilde d_3}$
are thereafter obtained from the general expression of \cite{Brezin}.\\
In the large-$x$ limit (corresponding to $T>T_c$ and small $H$), the expected
behaviour is given by Griffiths's analyticity condition \cite{Griffiths}
\be
f(x)\;=\;\sum_{n=1}^{\infty}a_n\,x^{\gamma-2(n-1)\beta}\,.
\label{Griffiths}
\ee
The form (\ref{eqstate}) of the equation of state is equivalent to the often
used relation
\be
M\;=\;h^{1/\delta} f_G(z),
\label{fG}
\ee
where $f_G$ is a further universal scaling function and $z$ the combination
\be
z\;=\;t'/h^{1/\beta\delta}.
\label{z}
\ee
The normalisation conditions of $f_G(z)$ are
\be
f_G(0)\; =\; 1~ \quad {\rm and}\quad f_G(z) 
\stackrel{z \rightarrow -\infty}{\longrightarrow}
(-z)^{\beta}~.
\label{normfg}
\ee
This version is normally used for comparison to QCD lattice data. The function
$f(x)$ is connected with $f_G(z)$ by
\be
y=f_G^{-\delta}, \quad x=z\,f_G^{-1/\beta}.
\label{connectFgandY}
\ee
These scaling functions are only valid close to $J_c$ and $H \to
0$. First tests show that the data we have used in the broken phase does not scale
directly, while in the high temperature phase most of the data scales close to $T_c$
and small $H$. So we used a more general form of (\ref{fG})
\be 
Mh^{-1/\delta}\; =\; \Psi (z, h^{\omega\nu_c})
\ee
with a scaling function $\Psi$, which can be expanded to
\ba
Mh^{-1/\delta}\;\!\!\!\!\!\!\!\!&=&\!\!\!\!\!\!\!\!\;f_G(z)\,+\,h^{\omega
  \nu_c}f_G^{(1)}(z)\nn\\
\!\!\!\!\!\!\!\!&&\!\!\!\!\!\!\!\!{}+\,h^{2\omega\nu_c}f_G^{(2)}(z)\,+\,\dots\,.
\label{fG_corr}
\ea
This way in the broken phase one obtains the leading part $f_G$ by
quadratic fits to our data in $h^{\omega\nu_c}$ at constant $z$-values and
different $(J/H)$-combinations. But we are only able to correct the
data with $z\ltapprox -2$ because we have not enough $J$-values closer to $J_c$
to make the fits. In Fig.\ \ref{fig:scafun} we show the influence of the
corrections and the final scaling function $f_G$ in the broken phase (dashed
\begin{figure}[ht]
   \epsfig{bbllx=83,bblly=214,bburx=496,bbury=587,
       file=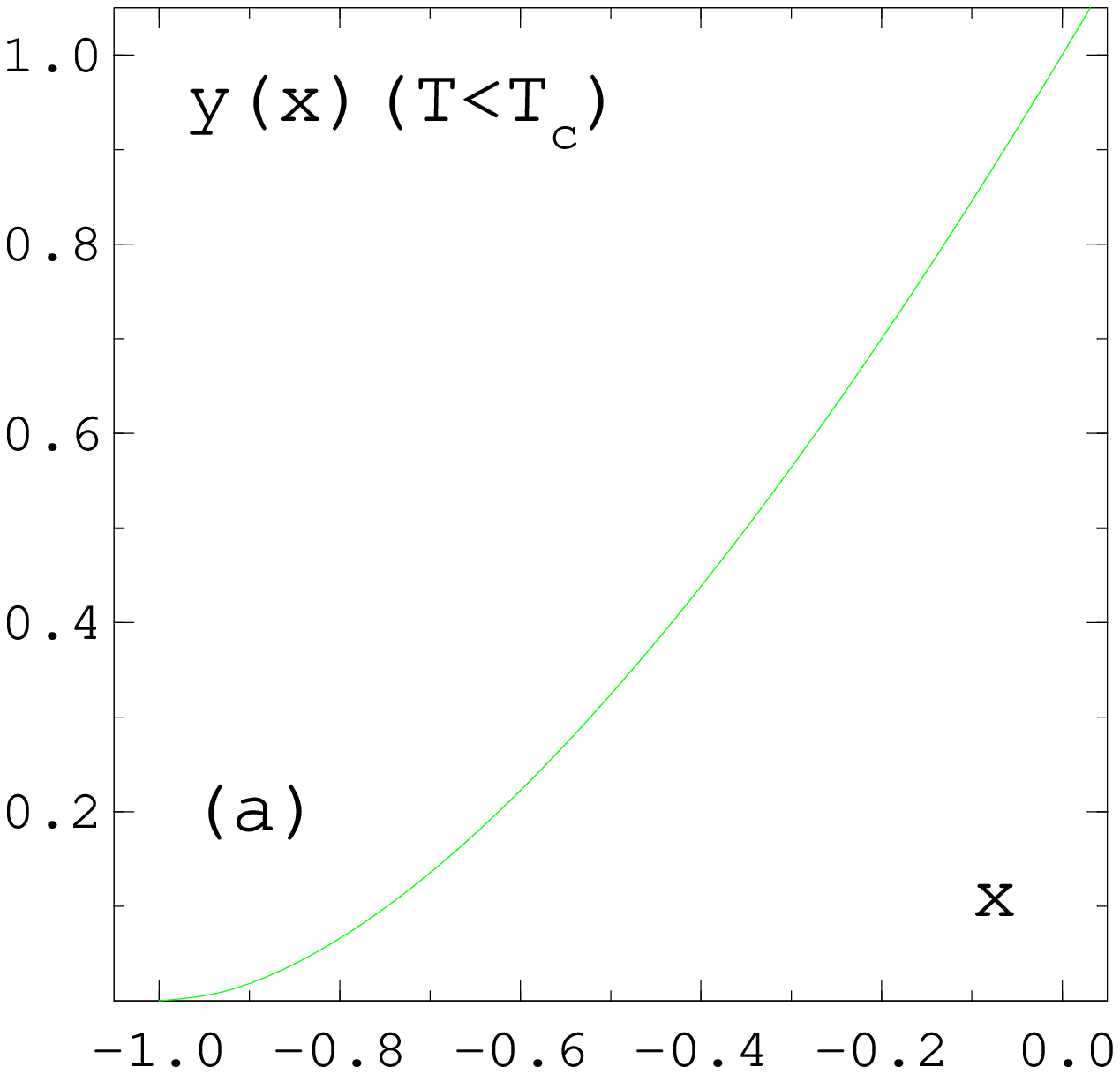, width=70mm}
   \epsfig{bbllx=83,bblly=264,bburx=496,bbury=587,
       file=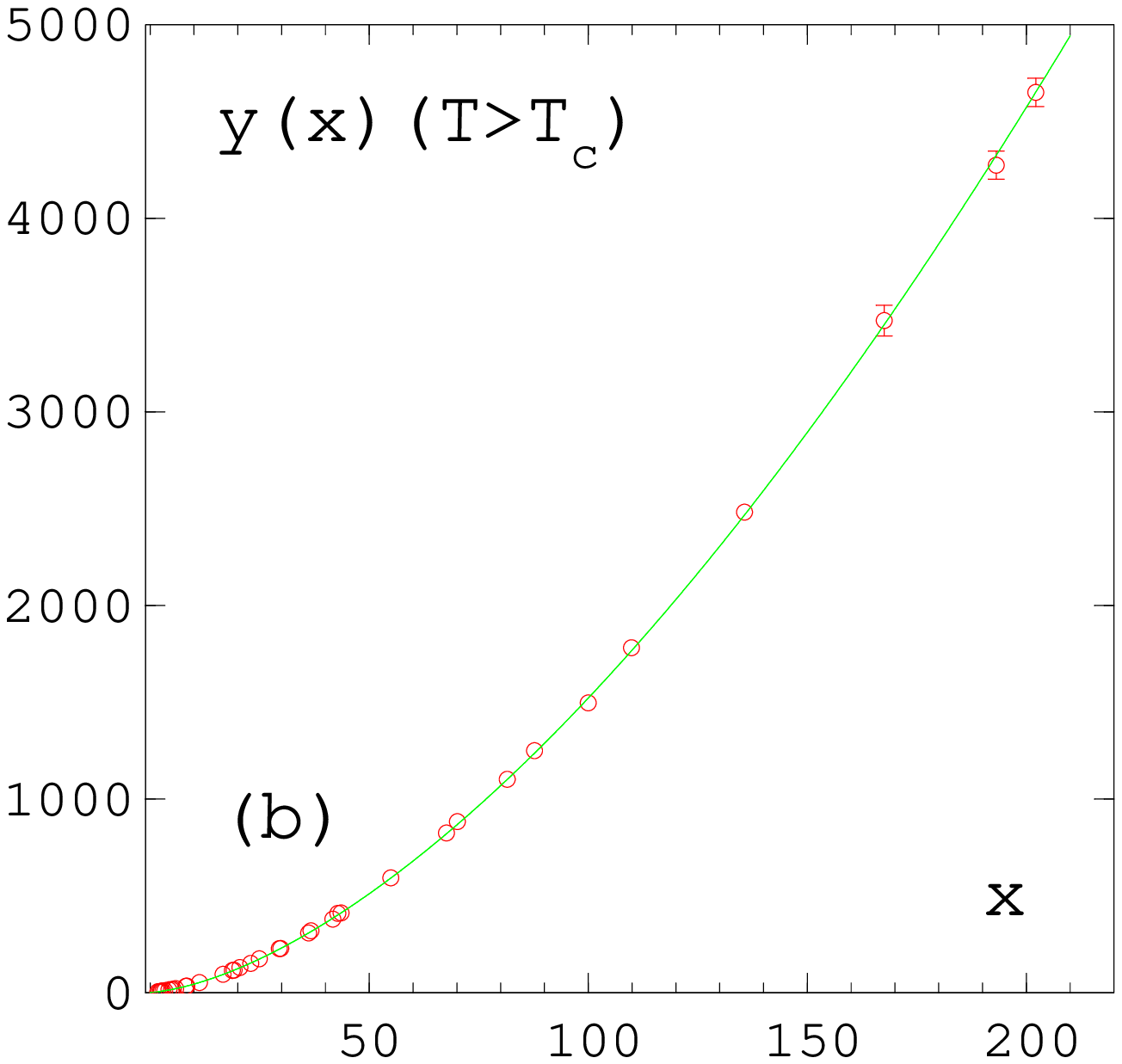, width=70mm}
\caption{The functions $y=f(x)$ at small $x$ (a) and large $x$-values (b). In
  figure (a) we plot the fit to ansatz (\ref{PTform}) using extrapolated
  data in the broken phase and data-points in the symmetric phase. In (b) we
  plot data-points of the symmetric phase and the fit to ansatz
  (\ref{Griffiths}) using the first three terms.}
\label{WGform}
\end{figure}
line).\\
Our result for $f_G(z)$ can be transformed with (\ref{connectFgandY})
into the Widom-Griffiths form of the equation of the state
(\ref{eqstate}). Unfortunately the $z$-interval we used to extract $f_G(z)$ is
only equivalent to the small region $-0.9\ltapprox x \ltapprox -0.7$, which can
be used for a fit. We use the three leading terms in (\ref{f_inv}) 
\ba
x_1(y)+1 \!\!\!\!\!\!\!\!&=&\!\!\!\!\!\!\!\! ({\widetilde c_1} \,+\, {\widetilde d_3})\,y \,+\,
             {\widetilde c_2}\,y^{1/2} \nn\\ 
             &&{}\!\!\!\!\!\!\!\!+\,{\widetilde d_2}\,y^{3/2} \;.
\label{PTform}
\ea
Since $y(0)=1$ the coefficients are connected by ${\widetilde
  d_2}=1-({\widetilde c_1} \,+\,{\widetilde d_3}\,+\,{\widetilde c_2})$. Fits
to $x$ in the interval $-0.9\ltapprox x \ltapprox -0.7$ and points in the
symmetric phase with $0.2\ltapprox x \ltapprox 2.9$, $1.4\ltapprox J
< J_c$ and $H \leq 0.0015$ lead to
\be
{\widetilde c_1} + {\widetilde d_3} \,=\, 0.36(5) \;,\quad
{\widetilde c_2} \,=\, 0.69(3).
\label{result}
\ee
\n The result of the fit is shown by the line in Fig.\ \ref{WGform}(a).\\
For large $x$ we use a 3-parameter fit of the first three terms of Griffiths's
analyticity condition (\ref{Griffiths})
\be
y_2(x)\;=\;a_1 x^{\gamma}\,+\,a_2 x^{\gamma-2\beta}\,+\,a_3 x^{\gamma-4\beta}
\ee
in the interval $x \in [1.75,202]$ and data points restricted to $1.4\ltapprox
J < J_c$ and $H \leq 0.0015$. We find
\ba
&&a_1\,=\, 0.92(1)\,, \quad  a_2 \,=\, 1.17(2)\,,\nn\\
&&\quad \quad \quad \quad a_3 \,=\,0.91(3).
\label{resulthighx}
\ea
This result is plotted in Fig.\ \ref{WGform}(b). With the coefficient $a_1$ of
the leading part one can calculate the universal ratio
\be
R_{\chi}\;=\;a_1^{-1}\;=\;1.09(1).
\ee
The $O(6)$ scaling function $f_G$ can be parametrically obtained from $x_1(y)$
\begin{figure}[t]
   \epsfig{bbllx=83,bblly=214,bburx=496,bbury=587,
       file=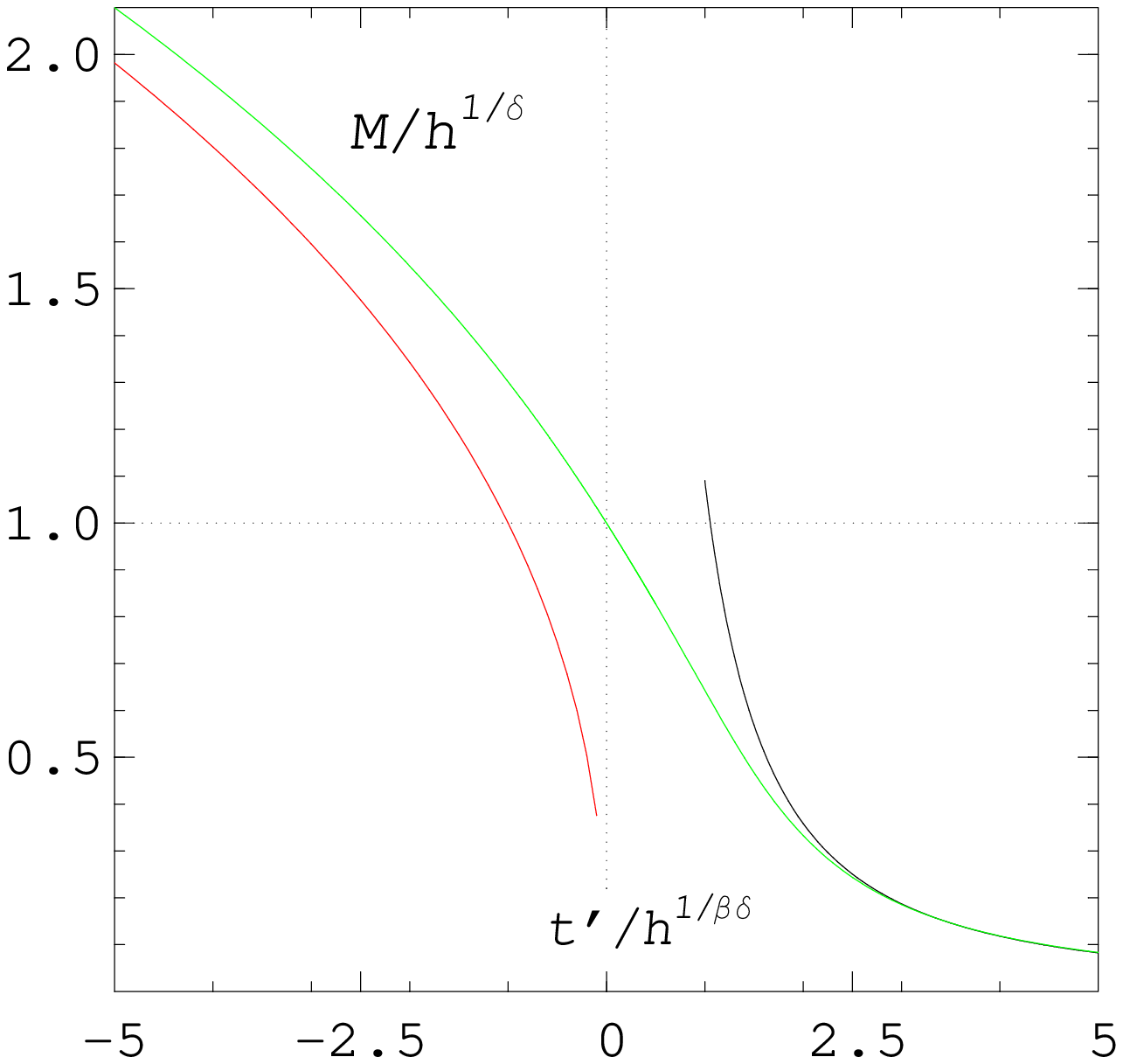, width=70mm}
   \epsfig{bbllx=83,bblly=264,bburx=496,bbury=587,
       file=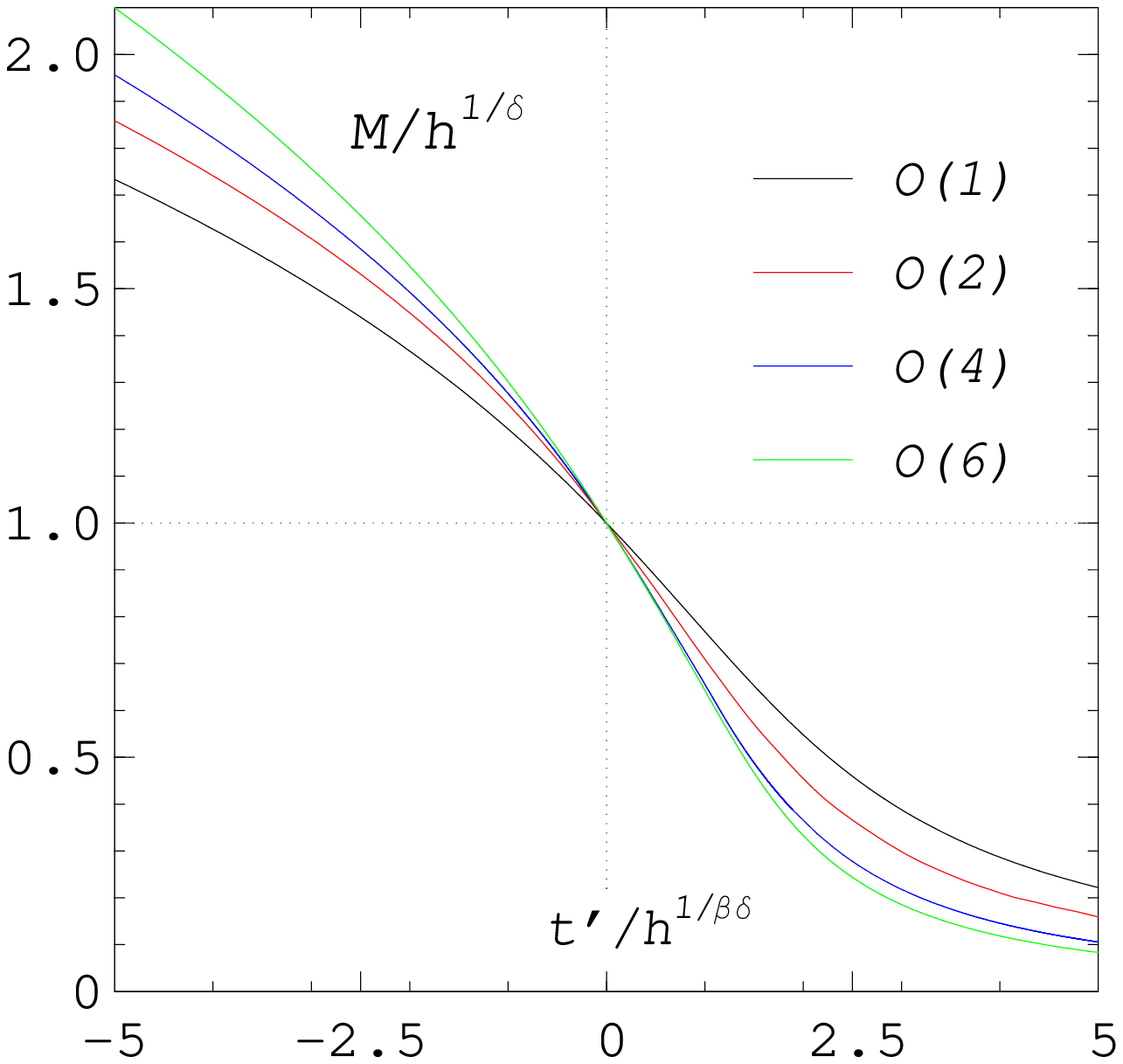, width=70mm}
\caption{(a) The scaling function $f_G$ of the O(6) model (green) and the
       asymptotic behaviours for $z\to \infty$ (black) and $z\to -\infty$
       (red). (b) The scaling function $f_G$ for the Ising (black), $O(2)$
       (red), $O(4)$ (blue) and $O(6)$ model (green).}
\label{allesfg}
\end{figure}
and $y_2(x)$, which is connected by a spline between $z=0.5$ and $z=0.8$, where
we have no reliable parametrisation. The result is plotted in Fig.\
\ref{allesfg}(a). Also plotted are the leading terms of the asymptotic behaviour at
$z\to\pm \infty$. These are
\ba
f_G (z)\!\!& \stackrel{z \rightarrow -\infty}{=}&\!\! (-z)^{\beta}~,\label{fgasym}\\
f_G (z)\!\!& \stackrel{z \rightarrow +\infty}{=}&\!\! R_{\chi} z^{-\gamma}
\label{fgasyp}
\ea
according to the normalisation (\ref{normfg}). The fact that for large
temperatures and small $H$ the magnetisation is proportional to $H$, see
Eq. (\ref{MpropH}), explains the asymptotic behaviour for $z\to \infty$. In the
symmetric phase the asymptotic behaviour is reached for small absolute values of $z$,
while in the broken phase the scaling function converges to the asymptotic form
not until large absolute values of $z$.\\
Finally, the $O(6)$ scaling function $f_G$ can be compared to the corresponding
functions for the Ising ($O(1)$) model \cite{Engels:2002fi}, the $O(2)$
\cite{Engels:2000xw} and the $O(4)$ model \cite{Engels:1999wf}, shown in Fig.\
\ref{allesfg}(b). All functions have a similar shape.

\section{The Pseudocritical Line}
\label{PCL}
\begin{figure}[t]
   \epsfig{bbllx=83,bblly=264,bburx=496,bbury=587,
       file=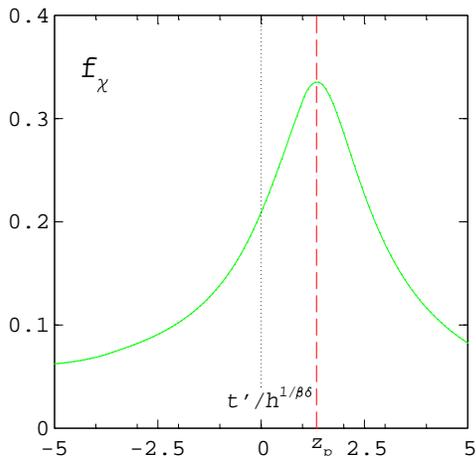, width=70mm}
\caption{The scaling function $f_\chi$ for the $O(6)$ model. The dashed red
       line shows the position of $z_p=1.34$.}
\label{fig:chil}
\end{figure}
\n In order to discuss finite-size-scaling functions in an easier way, it is
common to study lines of constant $z$-values. There one expresses $H$ as a
function of $T$ or vice versa. Important examples of lines of fixed $z$ are the
critical line ($z=0$), discussed in Section \ref{z=0}, and the pseudocritical line
$z=z_p=const$, the line of peak positions of the susceptibility $\chi_L$ in the
$(t,h)$-plane for $V\rightarrow \infty$. There are two different ways to find
that value of $z_p$ for $O(N)$ models. One way is to locate the peak positions
of $\chi_L$ as a function of the temperature at different fixed small values of
the magnetic field on lattices with increasing size $L^3$. The scaling function,
on the other hand, offers a more elegant way to determine the pseudocritical
\begin{figure}[ht]
   \epsfig{bbllx=83,bblly=214,bburx=496,bbury=587,
       file=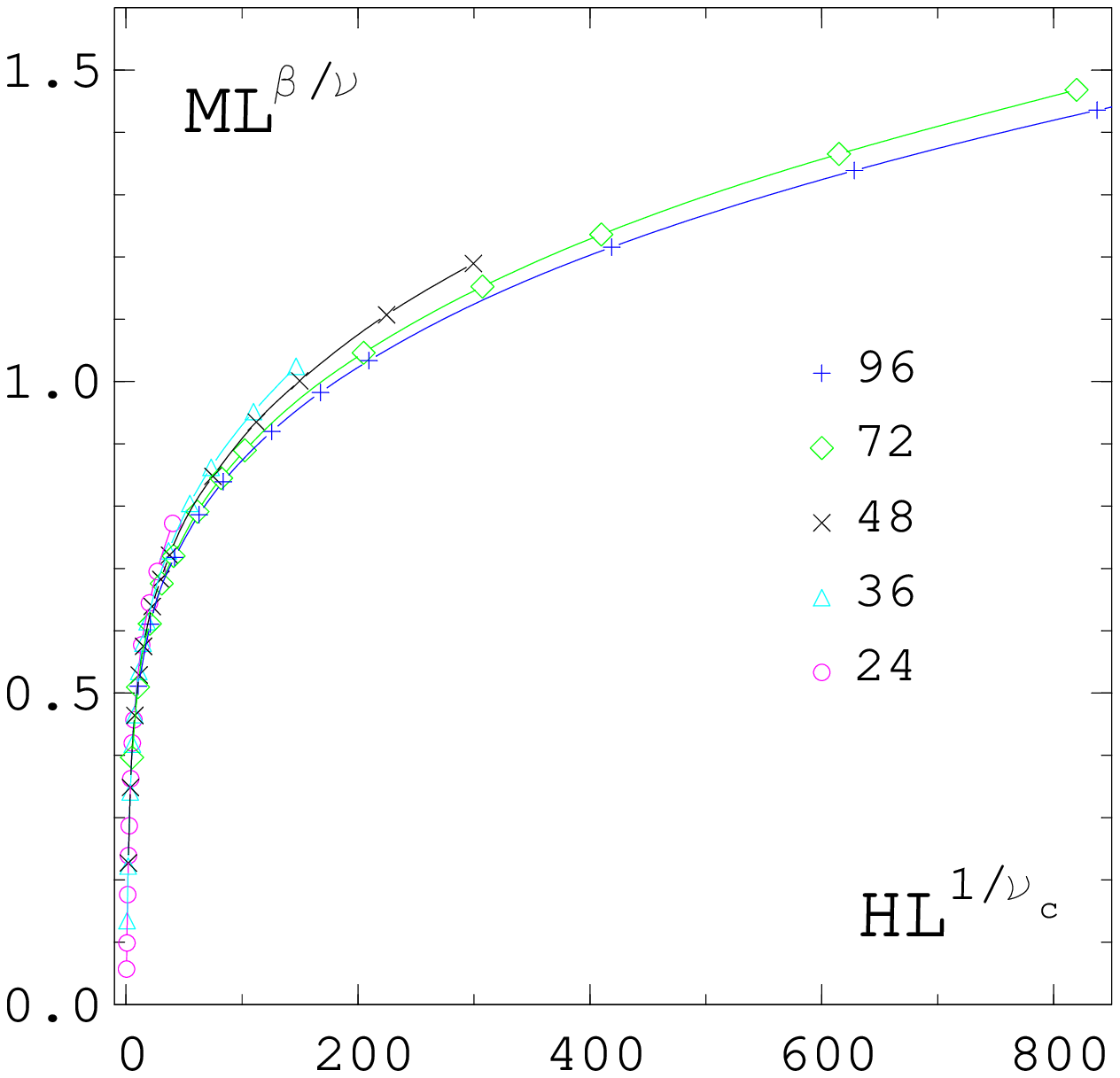, width=70mm}
   \epsfig{bbllx=83,bblly=264,bburx=496,bbury=587,
       file=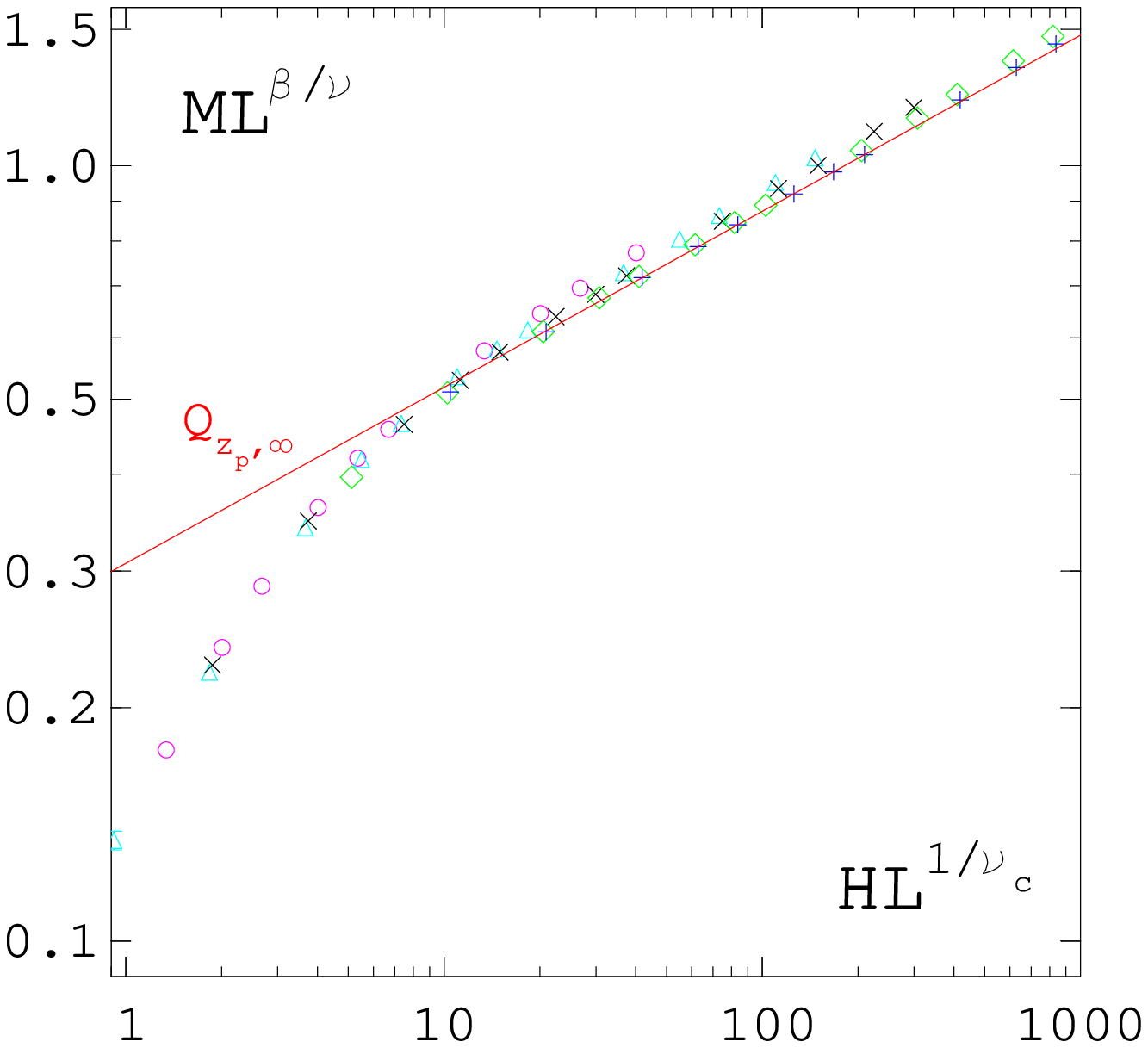, width=70mm}
\caption{(a) Finite-size scaling of $ML^{\beta/\nu}$ on the
   pseudocritical line. (b) is a
   double-log plot of (a). The solid line in (b) shows the asymptotic form
   $Q_{z_p,\infty}$, the symbols denote different lattice sizes $L$.}
\label{pseudoO6}
\end{figure}
\n line. Since $\chi_L$ is the derivative of $M$ 
\be
\chi_L={\partial M\over \partial H}={h^{1/\delta-1} \over H_0} f_\chi(z)~,
\label{max}
\ee
its scaling function $f_\chi(z)$ can be calculated directly from $f_G(z)$
\be
f_\chi(z)={1 \over \delta}
\left ( f_G(z) - {z\over \beta}\frac{\partial f_G}{\partial z}(z) \right)~.
\label{fgc}
\ee
The maximum of $f_\chi(z)$ is located at $z_p$, which is another universal
quantity. We find for the $O(6)$ model 
\be
z_p\;=\;1.34(5)\,.
\ee
The error includes the fit-errors of the parameters in (\ref{resulthighx}).\\
In Fig.\ \ref{fig:chil} we show the result for $f_\chi(z)$ from 
Eq.\ (\ref{fgc}) in the $O(6)$ model.\\
At $z_p$ a finite-size-scaling analysis in the variables $H$ and $L$ can be
performed. Eq. (\ref{mexpan}) reduces to
\be
M(H,L)\;=\; L^{-\beta/\nu} \, Q_{z_p} ( hL^{1/\nu_c})  \,+\, \dots
\label{mq0}
\ee
with another universal scaling function $Q_{z_p}$. The asymptotic form
$Q_{z_p,\infty}$ of $Q_{z_p}$ is
 \be
Q_{z_p} \stackrel{L\rightarrow \infty~}{\longrightarrow} Q_{z_p,\infty}=f_G(z_p)\,( hL^{1/\nu_c})^{1/\delta}
\quad
\label{Qasy}
\ee
The results are presented in Fig.\ \ref{pseudoO6}(a). The data does not scale directly
but with increasing volume the data-points approach $Q_{z_p}$ from the top. In
Fig.\ \ref{pseudoO6}(b) we plotted the data in a double-log form and find that
the asymptotic form $Q_{z_p,\infty}$ coincides with the $Q_{z_p}$-value of the
largest lattice extension at $HL^{1/\nu_c}\gtapprox 42$. Therefore $Q_{z_p}$
is asymptotic. At smaller values, one observes an approach of $Q_{z_p}$ from below
to $Q_{z_p,\infty}$.

\section{Conclusions}
\label{section:conclusion}
\n In this paper we calculated several important quantities of the O(6) spin
model directly from Monte Carlo simulations on cubic lattices. At zero external
field we determined the critical coupling $J_c$ by a finite-size-scaling
analysis of the Binder cumulant and by the $\chi^2$-method. Our result agrees
in the first four digits with the result of Butera and Comi. At the critical
point we estimated the critical exponents from finite-size-scaling fits. We
obtained $\beta$ from the magnetisation, $\gamma$ from the susceptibility
and $\nu$ from the derivative of the Binder cumulant. Our results are in accord
with the values found by Butera and Comi but slightly different compared to the
values found by Antonenko and Sokolov. We find small corrections to scaling
for all observables.\\
On the critical line $T=T_c,H>0$ and in the limit $V \to \infty$, the critical
amplitude $d_c$ of the magnetisation was computed. We found small
negative corrections to scaling and checked the finite-size-scaling
behaviour of $M$ at $T_c$ and its asymptotic form.\\
Below the critical temperature, we investigated the behaviour of $M$ at several
couplings $J$ as a function of $H^{1/2}$ in the limit $V \to \infty$. Close to
the coexistence line, i.e. small $H\to 0$, the predicted Goldstone behaviour
was observed. We were able to extrapolate our data to the values $M(T<T_c;H=0)$ of
the infinite volume limit, fitted these $M$-values with the corresponding
ansatz, and estimated the critical amplitude $B$ of the magnetisation. In this
case the corrections to scaling were again negative and more pronounced as
on the critical line. At high temperatures and $H>0$, we observe the expected
proportional dependence on $H$ of the magnetisation.\\
We used our data of the largest lattices in the low and high-temperature phase to
parametrise the scaling function $f_G$ of the O(6) model. We encountered
large corrections to scaling in the broken phase, while most data in the
symmetric phase scales directly. By generalizing $f_G$ to include corrections to
scaling, our group extracted a part of $f_G$ in the broken phase and fitted the result
combined with direct data points in the symmetric phase. On the other hand, we
fitted data of the symmetric phase using Griffiths's analyticity
condition. Finally, we compared our $O(6)$-result for $f_G$ with the
corresponding scaling functions of the $O(1)$, $O(2)$ and $O(4)$ model. These
functions are clearly distinguishable and in a systematic order. We use our
result of $f_G$ to calculate the scaling function $f_\chi$ of the
susceptibility. From the position of the maximum in $f_\chi$ the location of the
pseudocritical line was determined. There we made finite-size-scaling plots and
found considerable corrections to scaling. The data of smaller lattices
approaches the universal finite-size-scaling function from above. The asymptotic
form of the universal part is reached at $HL^{1/{\nu_c}} \approx 42$.\\ 
A comparison between the $O(6)$ model and aQCD will be done in the near
future.\\
   
\begin{acknowledgments}
\n We are grateful to J\"urgen Engels and Sandra Wanning for reading this
manuscript carefully. Our work was supported by the Deutsche
Forschungs\-ge\-meinschaft under Grant No.\ FOR 339/1-2.
\end{acknowledgments}


\clearpage

\end{document}